\documentclass{article}

\usepackage{booktabs} % For formal tables

\usepackage{geometry}
\usepackage{algorithmicx}
\usepackage[ruled]{algorithm}
\usepackage[noend]{algpseudocode}
\algrenewcommand\algorithmicindent{0.70em}%

\usepackage{graphicx}
\usepackage{subcaption}
\usepackage{amsmath, mathtools, calc}
\usepackage{amsmath,amsthm, amsfonts,amssymb,epsfig,epstopdf}

\usepackage{amsfonts}
\usepackage{url}
\usepackage{array}[2003/12/17]
\usepackage{calc}

\newlength\celldim \newlength\fontheight \newlength\extraheight
\newcounter{sqcolumns}
\newcolumntype{S}{ @{}
>{\centering \rule[-0.5\extraheight]{0pt}{\fontheight + \extraheight}}
p{\celldim} @{} }
\newcolumntype{Z}{ @{} >{\centering} p{\celldim} @{} }
{\end{tabular}}

\usepackage{xparse}
\begin{document}

\title{Fast Strassen-based $A^t A$ Parallel Multiplication}

%\titlenote{Produces the permission block, and copyright information}

\author{Viviana Arrigoni, Annalisa Massini\\
Computer Science Department \\
   Sapienza University of Rome, Italy\\
   \texttt{\{arrigoni,massini\}@di.uniroma1.it }
   }
%\authornote{....}
%\orcid{1234-5678-9012}
%\affiliation{%
%  \institution{Computer Science Dept. -
%   Sapienza, Univ. of Rome}
%  \streetaddress{Via Salaria, 113}
%  \city{Rome}
%  \state{Italy}
%  \postcode{43017-6221}
%}

%\author{Annalisa Massini}
%\authornote{The secretary disavows any knowledge of this author's actions.}
%\affiliation{%
%  \institution{Computer Science Dept. - 
%   Sapienza, Univ. of Rome}
%  \streetaddress{Via Salaria, 113}
%  \city{Rome}
%  \state{Italy}
%  \postcode{43017-6221}
%}
%\email{massini@di.uniroma1.it}

%\title{Strassen based}
%\date{}

%\begin{document}

\maketitle

\begin{abstract}

Matrix multiplication $A^t A$ appears as intermediate operation during the solution of a wide set of problems.
In this paper, we propose a new cache-oblivious algorithm for the $A^t A$ multiplication. Our algorithm, \normalsize{A}\footnotesize{T}\normalsize{A}, calls classical Strassen's algorithm as sub-routine, decreasing the computational cost %(expressed in number of performed products) 
of the conventional $A^t A$ multiplication to $\frac{2}{7}n^{\log_2 7}$.
It works for generic rectangular matrices and exploits the peculiar symmetry of the resulting product matrix for sparing memory. We used the MPI paradigm to implement \normalsize{A}\footnotesize{T}\normalsize{A} in parallel, and we tested its performances on a small subset of nodes of the Galileo cluster. Experiments highlight good scalability and speed-up, also thanks to minimal number of exchanged messages in the designed communication system. Parallel overhead and inherently sequential time fraction are negligible in the tested configurations.
\end{abstract}
%\keywords{Matrix Multiplication, Parallel Algorithms, MPI}

%\maketitle
%\end{document}
\section{Introduction}
	
Matrix multiplication is probably the main pillar of linear algebra. It is a fundamental operation in many problems of mathematics, physics, engineering, and computer science. 
The algorithmic aspects of matrix multiplication have been extensively explored, as well as its parallelization.	Many approaches have been used over the years. In particular distributed computing and high-performance computing have been taken into considerations with the aim of optimizing performance and obtaining more and more efficient parallel algorithms and implementations.

Matrix $A^t A$ is a particular matrix multiplication  involved in several applications. %used for solving many problems
In Strang's book \cite{strang06}, the product $A^t A$ is extensively used for its many useful properties.
In fact, matrix $A^t A$ is symmetric and positive-definite. Product $A^t A$ appears as intermediate operation in many  situations. For example, it appears in the projection matrix $P = A(A^t A)^{-1} A^t$, and in the equations $A^t A \hat{x} = A^t b$, known as the normal equations.  Furthermore, it is used in the least square problem, in the Gram-Schmidt orthogonalization process, in the Singular Value Decomposition, and has many other applications (see, e.g., \cite{strang06}).
	
In this work, we propose a new parallel algorithm to compute the product $A^t A$. Our algorithm is based on the Strassen’s strategy for the fast matrix multiplication. 
Strassen’s algorithm \cite{strassen1969gaussian} is the most used fast algorithm for matrix multiplication. 
In fact, Strassen broke the $O(n^3)$ operation count for executing the product among two matrices, reorganizing the recursive matrix multiplication algorithm, that is replacing a multiplication step with 18 cheaper matrix additions.
This substitution implies asymptotically fewer multiplications and additions, and provides an algorithm executing $O(n^{2.81})$ operations. 
Winograd's variant improves Strassen’s complexity by a constant factor replacing one matrix multiplication with 15 matrix additions.
%It is cache oblivious, and meets the lower bounds with respect to memory requirements, computational costs, and communication costs.
	
Since our algorithm exploits the characteristics of the resulting product matrix, it results to be faster than other parallel matrix multiplication algorithms, such as those based on classical $\Theta(n^3)$ multiplication, and those based on  Strassen-like matrix multiplications. We study the performance of our algorithm by performing a set of tests on matrices of size $5^3$ and 10$^4$. The MPI paradigm has been used to develop the parallel implementation, and realize tests running on a multiprocessor system.
%and compare it with a very fast parallel algorithm for matrix multiplication, that is CAPS \cite{}.
	
The paper is organized as follows. Section~\ref{sec:related} illustrates the related work. Section~\ref{sec:algo} describes the proposed algorithm in its sequential version, whilst Section~\ref{sec: par_imp} gives the description of the parallel algorithm, as well as details on the parallel implementation.
In Section~\ref{sec:comm-cost} considerations on the communication costs are provided.
Section~\ref{sec:performance} illustrates the results of the experimental phase, obtained using several performance assessment parameters, and discuss their behaviour.
Finally, Section~\ref{sec:conclusions} summarizes the characteristics of the proposed algorithm and outlines the ideas for future work. 

\section{Related work}
\label{sec:related}

%\subsection{Strassen}

Since Strassen’s algorithm proposal, many fast matrix multiplication algorithms were designed and improved the asymptotic complexity (see, e.g., \cite{LeGall-SAC14,Stothers-03,Williams-STOC12} for recent algorithm). Unfortunately, often improvements come at the cost of very large hidden constants.
Since for small matrices, Strassen’s algorithm has a significant overhead, several authors have designed hybrid algorithms, deploying Strassen’s multiplication in conjunction with conventional matrix multiplication, see, e.g., \cite{brent-TR70, brent-NM70, Higham-TOMS90, Huss-Super96, Benson-PPOPP15}.

%\subsection{HASA}
In \cite{d2007adaptive}, Authors extend Strassen’s algorithm to deal with rectangular and arbitrary-size matrices.
They consider the performance effects of Strassen’s
directly applied to rectangular matrices or, after a cache-oblivious problem division, to (almost) square matrices, thus exploiting data locality. 
They also exploit the state-of-the-art adaptive software packages ATLAS and hand tuned packages such as GotoBLAS. Besides, they show that choosing a suitable combination of Strassen’s with ATLAS/GotoBLAS, their approach achieves up to 30\%-22\% speed-up versus ATLAS/GotoBLAS alone on modern high-performance single
processors.
%%%%%%%%%%%%%%%%%%%%%%%%

%\subsection{Parallel Algorithms for Strassen’s Matrix Multiplication} 
Numerous parallel algorithms for Strassen’s Matrix Multiplication have been proposed.
In \cite{Luo-SAC95}, Luo and Drake explored Strassen-based parallel algorithms that use the communication patterns known for
classical matrix multiplication. 
They considered using a classical 2D parallel algorithm and using Strassen locally.  They also considered using Strassen at the highest level and performing a classical parallel algorithm for each sub-problem generated, where the size of the sub-problems depends on the number of Strassen steps taken. Further, the communication costs for the two approaches is also analyzed.
In \cite{Grayson-PPL95}, the above approach is improved using a more efficient parallel matrix multiplication algorithm and running on a more communication-efficient machine. 
They obtained better performance results compared to a purely classical algorithm for up to three levels of Strassen's recursion.
In \cite{Kumar-IPPS93}, Strassen's algorithm is implemented on a shared-memory machine. The trade-off between available parallelism and total memory footprint is found by differentiating between \emph{partial} and \emph{complete} evaluation of the algorithm.
%corresponding  to what we call depth-first and breadth-first traversal of the recursion tree. 
The Authors show that by using $\ell$ \emph{partial} steps before using \emph{complete} steps, the memory footprint is reduced by a factor of $(7/4)^{\ell}$ compared to using all \emph{complete} steps. 
%They did not consider communication costs in their work.
Other parallel approaches \cite{Desprez-04, Hunold-08, Song-PDCS06} have used more complex parallel schemes and communication patterns, but consider at most two steps of Strassen and obtain modest performance improvements over classical algorithms.

%\subsection{Communication-Optimal Parallel Algorithm for Strassen’s Matrix Multiplication}
In \cite{ballard2012communication}, a parallel algorithm based on Strassen's fast matrix multiplication, Communication-Avoiding Parallel Strassen (CAPS), is described.
Authors present the computational and communication cost analyses of the algorithm, and show that it matches the communication lower bounds described in \cite{BDHS-SPAA11}.

In this work, we consider a particular matrix multiplication, that is the multiplication between $A^t$ and $A$, where $A$ may have any size and shape. 
We exploit the recursive Strassen's algorithm, that is recursively applied to conceivably rectangular matrices,
%and in the different recursive steps we apply Strassen, both to square and rectangular matrices, 
exploiting %for the latter 
the idea described in \cite{d2007adaptive}. 
%The naive parallel implementation, proposed in this paper is similar to that described in \cite{ballard2012communication}.

To the best of our knowledge, this is the first parallel algorithm specifically thought for computing the product $A^t A$.

\section{Algorithm for $A^t A$}
\label{sec:algo}

In this section, we describe the recursive algorithm for the matrix multiplication $A^t \cdot A$, denoted as A\footnotesize{T}\normalsize{A} algorithm. 

The algorithm works for general $m\times n$ rectangular matrices. 
At each recursive step, the matrix $A$ is divided in four sub-matrices as: 
\vspace{-0.5cm}
\begin{center}
\large
\[  A=
\begin{bmatrix}
\label{eq: submatrices}
       A_{1,1} & A_{1,2}  \\
A_{2,1} & A_{2,2}
     \end{bmatrix}
\]
\end{center} 
If we define: $m_1 \overset{def}{=}\left \lceil \frac{m}{2}\right \rceil$, $m_2  \overset{def}{=}\left \lfloor \frac{m}{2}\right \rfloor$, $n_1 \overset{def}{=}\left \lceil \frac{n}{2}\right \rceil$, $n_2 \overset{def}{=}\left \lfloor \frac{n}{2}\right \rfloor$, then, for the four sub-matrices composing $A$ we have: 
\noindent
\begin{equation}
\begin{aligned}
A_{1,1} \in \mathbb{R}^{m_1 \times n_1},\\
A_{1,2}\in \mathbb{R}^{m_1 \times n_2},\\
A_{2,1}\in \mathbb{R}^{m_2 \times n_1},\\
A_{2,2}\in \mathbb{R}^{m_2 \times n_2}. 
\end{aligned}
\end{equation}

Also the product matrix $C = A^t \cdot A$ can in turn  be divided into four sub-matrices. The four components of $C$ are obtained using the four components of the matrices $A$ and $A^t$ for executing the product. Thus, $C$ consists of the following four sub-matrices: %can be written as:
%\begin{center}
%    $A^t \cdot A=$ 
%    \large
%    \begin{squarecells}{2}
%    $C_{1,1}$ & $C_{1,2}$ \nl
%    $C_{2,1}$ & $C_{2,2}$ \nl
    
%    \end{squarecells}
%\normalsize
%\begin{squarecells}{2}
%$A_{1,1}^tA_{1,1} + A_{2,1}^tA_{2,1}$ & $A_{1,1}^tA_{1,2} + A_{2,1}^tA_{2,2}$  \nl
%$A_{1,2}^tA_{1,1} + A_{2,2}^tA_{2,1}$ & $A_{1,2}^tA_{1,2} + A_{2,2}^tA_{2,2}$ \nl

%\end{squarecells}
%\end{center}
\noindent
\begin{equation}
\begin{aligned}
\label{eq: Cij}
 C_{1,1} = A_{1,1}^tA_{1,1} + A_{2,1}^tA_{2,1}\in \mathbb{R}^{n_1 \times n_1},\\
C_{1,2} = A_{1,1}^tA_{1,2} + A_{2,1}^tA_{2,2}\in \mathbb{R}^{n_1 \times n_2},\\
C_{2,1} = A_{1,2}^tA_{1,1} + A_{2,2}^tA_{2,1}\in \mathbb{R}^{n_2 \times n_1},\\
C_{2,2} = A_{1,2}^tA_{1,2} + A_{2,2}^tA_{2,2}\in \mathbb{R}^{n_2 \times n_2}.
\end{aligned}
\end{equation}
Both $C_{1,1}$ and $C_{2,2}$ components of matrix $C$ consist of two addends that are, in their turn, the left hand product of a matrix by its transpose, that is a product of type $M^T M$. 
Hence, four recursive calls are employed to compute the  sub-products $A_{1,1}^t A_{1,1}$ and $A_{2,1}^t A_{2,1}$ to obtain $C_{1,1}$, and $A_{1,2}^t A_{1,2}$ and $A_{2,2}^tA_{2,2}$ to obtain $C_{2,2}$. 

Since for any matrix $A$ the product $A^t\cdot A$ is symmetric, at each recursive step only the lower triangular part of the product matrix is stored. This allows to spare almost half of the memory occupation with respect to a full matrix representation, that is $n(n + 1)/2$ entries versus the usual $n^2$. For this reason, the term $C_{1,2}$  is not returned by the algorithm (being $C_{1,2} = C_{2,1}^t$). 
As for component $C_{2,1}$, in order to compute $A_{1,2}^t A_{1,1}$ and $A_{2,2}^tA_{2,1}$, we implemented the generalized Strassen's algorithm for non-square matrices presented in \cite{d2007adaptive}, denoted as HASA.
Finally, matrices $A_{1,2}$ and $A_{2,2}$ are transposed using the cache oblivious algorithm for matrix transposition shown in \cite{kumar2003cache}.

In Algorithm \ref{alg: AtA} the pseudo-code of the A\footnotesize{T}\normalsize{A} algorithm is provided. The base-case occurs as the number of rows or of columns of the current sub-matrix is less than or equal to 32. This size has been chosen taking into account several considerations, including experimental tests and  observations highlighted in \cite{douglas1994gemmmw} on the cost difference between performing an arithmetic operation and loading/storing operations. %to the Cache Line Size (CLS). 
In that case, the multiplication is performed using a non-recursive algorithm for matrix multiplication. The initialization of $m_i$, $n_j$, and $A_{i,j}$, $i,j = 1,2$, is performed as  described above. 
%in this section. 

\begin{algorithm}
\label{alg: AtA}
\caption{A\footnotesize{T}\normalsize{A} - Serial}
\label{alg: AtA}
\textbf{Input:} $A\in \mathbb{R}^{m\times n}$ \\
\textbf{Output:} Lower triangular part of $C = A^t \cdot A$
\begin{algorithmic}[1]

\Procedure{AtA}{$A$, $m$, $n$}

\If {$m \lor n <=$ 32}  \\  \qquad \Return mult($A$, $m$, $n$);
\Else
\State Define $m_1$, $n_1$, $m_2$, $n_2$;
\State Initialize $A_{i,j}$, $i,j = 1,2$;
\State $S_1\leftarrow$ A\scriptsize{T}\normalsize{A}($A_{1,1}$, $m_1$, $n_1$); 
\State $S_2\leftarrow$ A\scriptsize{T}\normalsize{A}($A_{2,1}$, $m_2$, $n_1$); 
\State $S_3 \leftarrow$ A\scriptsize{T}\normalsize{A}($A_{1,2}$, $m_1$, $n_2$); 
\State $S_4\leftarrow$ A\scriptsize{T}\normalsize{A}($A_{2,2}$, $m_2$, $n_2$); 
\State $S_5\leftarrow$ HASA($A_{1,2}^t$, $A_{1,1}$, $n_2$, $m_1$, $n_1$); 
\State $S_6\leftarrow$ HASA($A_{2,2}^t$, $A_{2,1}$, $n_2$, $m_2$, $n_1$); 
\State \Return 
\begin{center}
\[ C = 
\begin{bmatrix}

S_1 + S_2 &   \\
S_5 + S_6 & S_3 + S_4 \\

\end{bmatrix} 
\]

\end{center}
\EndIf
\EndProcedure
\end{algorithmic}
\end{algorithm}

%\subsubsection{HASA}
%In order to perform generic matrix multiplications $A_{1,2}^tA_{1,1}$ and $A_{2,2}^tA_{2,1}$ we employed Strassen's algorithm for general matrices, that is a generalization of the original Strassen's algorithm for non-square matrices and whose size is not a power of 2. 

As anticipated, we implemented the pseudocode shown in \cite{d2007adaptive} for matrix multiplications $A_{1,2}^tA_{1,1}$ and $A_{2,2}^tA_{2,1}$, and we used the same notation; we refer to such procedure as to HASA (originally standing for Hybrid ATLAS/GotoBLAS–Strassen algorithm), keeping the same name used in \cite{d2007adaptive} for a matter of reference and convenience, although we do not use ATLAS and GotoBLAS packages.

\subsection{Computational cost}
Strassen's algorithm is a cache oblivious algorithm to compute the product of two matrices and it was first described in \cite{strassen1969gaussian}. It achieves to perform a $2\times 2$ matrix multiplication using 7 multiplications instead of 8. By recursion, the general $n\times n$ matrix multiplication is performed using $O(n^{\log_2 7})$ multiplications. In Algorithm \ref{alg: AtA}, there are four recursive calls to A\footnotesize{T}\normalsize{A} on basically halved dimensions, and two calls to HASA. Thus we can derive the general recursive function: 
\begin{equation}
    T(n) = 4T\left (\frac{n}{2}\right ) +\frac{2}{7} n^{\log_2 7}.
\end{equation}
Using the Master Theorem (see, e.g., \cite{CLR09}), it holds that the number of multiplications performed by A\footnotesize{T}\normalsize{A} is upper-bounded by $\frac{2}{7}n^{\log_2 7}$.

\section{Parallel implementation}
\label{sec: par_imp}
The recursive sequential Algorithm \ref{alg: AtA} has been implemented in parallel for a multiprocessor system using MPI (Message Passing Interface).  MPI  is a message-passing library specification that provides a powerful, efficient and portable way to write parallel algorithms, \cite{nielsen2016introduction}. 

The parallelization idea consists in executing  the recursive calls to A\footnotesize{T}\normalsize{A} and to HASA (lines 7 to 12 in the sequential Algorithm \ref{alg: AtA}) in a concurrent fashion.
%Since HASA-P is recalled twice for each execution of
A parallel multiprocessor implementation has been developed for both  A\footnotesize{T}\normalsize{A} and HASA algorithms. We shall refer to these parallel routines as to A\footnotesize{T}\normalsize{A}-P and HASA-P, where $P$ denotes the number of available processors. 
Depending on the value of $P$, a certain number of parallel levels can be executed. A parallel level is a parallel execution of A\footnotesize{T}\normalsize{A}-P and of HASA-P, where at least two processors share out the recursive calls. 
Each parallel execution of A\footnotesize{T}\normalsize{A}-P requires at most six processes, one for each of the four calls to A\footnotesize{T}\normalsize{A}-P and the two calls to HASA-P.
Instead, each parallel execution of HASA-P requires at most seven processes, since HASA implements the Strassen's algorithm and executed seven recursive calls. 
When the maximum number of parallel levels is reached, processes execute sequentially A\footnotesize{T}\normalsize{A} or HASA, depending on the task that has been assigned to them. 

During the parallel phase all processes work independently from one another and do not need to interact nor to exchange data. Communication is resumed at the end of the sequential phase that is executed at the bottom of the recursive tree, when the output arguments of the recursive calls are collected and arranged to form the product matrix $C = A^t \cdot A$ on each parallel level.
%Observe that, thanks to the choice of $m_i$ and $n_j$, $i, j = 1,2$, described in the previous section, the workload is distributed among processes of a parallel level in the most even way. \\
In the following subsections, we describe in detail how A\footnotesize{T}\normalsize{A}-P and HASA-P have been implemented and the structures created in order to manage cooperation between processes. We shall also explain how the communication system was implemented.

\subsection{A\small{T}\Large{A}-P insights} 
\label{sec: ATAins}
In the previous section, we introduced the notion of parallel levels; the number of parallel levels is the maximum number of parallel executions of A\footnotesize{T}\normalsize{A}-P and HASA-P where at least two processes are responsible for the recursive calls.

We shall classify parallel levels as either \emph{complete} or \emph{incomplete}. 
A  parallel level is \emph{complete} when six processes are assigned to each call to A\footnotesize{T}\normalsize{A}-P and seven processes are assigned to each call to HASA-P. 
%where each of the six calls to functions is assigned to a different process, or to  an  execution of  where the seven recursive calls are assigned to seven distinct . 
On the contrary, a parallel level is \emph{incomplete} when the number of processes to which recursive calls are assigned in a recursive step of A\footnotesize{T}\normalsize{A}-P or HASA-P is less than six and seven, respectively.

\begin{figure}[h]
  \centering
  \includegraphics[width=0.3\textwidth]{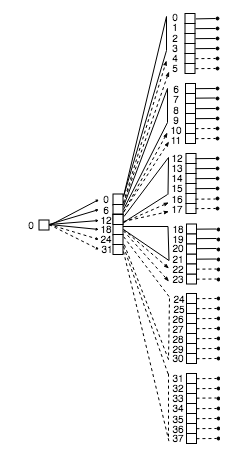}
\caption{\small{Representation of how processes are distributed among two complete parallel levels. Squares represent processes, and the number next to them are their identity ranks. Solid and dotted lines between processes represent recursive calls to A\scriptsize{T}\small{A}-P and HASA-P, respectively. After the second complete parallel level, each process executes either A\scriptsize{T}\small{A} or HASA independently.}}
\label{fig: comparlev}
\end{figure}

Every time  A\footnotesize{T}\normalsize{A}-P is called recursively in a complete parallel level, four processes will call  A\footnotesize{T}\normalsize{A}-P on the four sub-matrices of $A$ defined in Equation (\ref{eq: submatrices}), whilst the other two processes will execute HASA-P to compute the two rectangular products that are required to obtain $C_{2,1}$ (see equation (\ref{eq: Cij})). In a complete parallel level, HASA-P execution implies that seven processes execute HASA-P, each on its own sub-data. A representation of how processes are distributed in A\footnotesize{T}\normalsize{A}-P is depicted in Figure \ref{fig: comparlev}, that shows the tree structure of the processes organization for two complete parallel levels.  

The structure described above, according to which A\footnotesize{T}\normalsize{A}-P or HASA-P are called, can be used to derive the function $npl(\ell)$, that we use to calculate the number of processes needed to accomplish $\ell$ complete parallel levels of  A\footnotesize{T}\normalsize{A}-P. In particular it holds that:
\noindent
\begin{equation}
    \begin{aligned}
    npl(0) = 1 \quad npl(1) = 6,\quad \qquad\\
    npl(\ell > 1) = 6\cdot4^{\ell -1} + 2 \sum\limits_{k = 0}^{\ell - 2}4^k \cdot 7^{\ell -1 -k}.
    \end{aligned}
\end{equation}

Given the number of available processors $P$, the maximum number of executable complete parallel levels  $\ell_{max}$ is:
\begin{equation}
\label{eq: lmax}
    \ell_{max} = \max \left \{\ell \,| \, npl(\ell) \leq P \right \}.
\end{equation}
If the number of available processes $P$ is equal to $npl(\ell)$, for some  $\ell$, then no more processes are available and therefore all active processes will continue their task sequentially. 

The alternative case occurs when $npl(\ell_{max}) < P < npl(\ell_{max} + 1)$. This corresponds to the scenario where no more complete parallel levels can be executed, yet some computational resources are still available. In this case the remaining $P - npl(\ell_{max})$ processes are distributed in order to lighten the workloads of those processes involved in the complete parallel levels. When this event occurs, an incomplete parallel level is set between the $\ell_{max}$-th complete parallel level and the sequential calls.  

When an incomplete parallel level can be performed, processes are distributed according to the following idea. 
To spread the $P - npl(\ell_{max})$ processes as evenly as possible, the value $k$ is calculated as follows: 
\begin{align}
\begin{aligned}
\label{eq: k}
k = \max \big\{ &k |\, k\cdot npl(\ell_{max}) \leq \\
&\leq P - npl(\ell_{max}) < \\
&<(k + 1)\cdot npl(\ell_{max})\big\}.
\end{aligned}
\end{align}
Then, $k$ of the still available processes are paired to each process involved in the last complete parallel level. If $P - npl(\ell_{max})> k\cdot npl(\ell_{max})$, there are other $P - (k + 1) \cdot npl(\ell_{max})$ processes to distribute.
The distribution is realized following a hierarchical chart that exploits a classification  of already active processes, based on  the heaviness of the task they have to work on. 
First, we can observe that HASA-P is computationally more expensive than A\footnotesize{T}\normalsize{A}-P, since it involves seven recursive calls instead of six. Hence HASA-P has the highest priority.
The second  factor used to establish process priority is represented by the size of the sub-problem assigned to a call. 
In fact, if $m$ and $n$ are not a power of 2, after a certain number of recursive calls, it holds that $m_1=m_2 + 1$ and $n_1 = n_2 + 1$. Therefore, the size of the data of some processes may be higher.

\begin{figure}[h]
  \centering
  \includegraphics[width=0.25\textwidth]{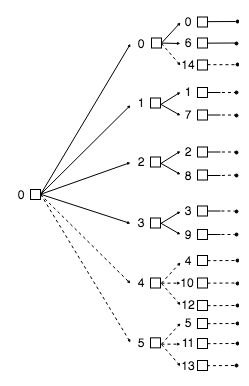}
\caption{\small{Representation of how processes are distributed among one complete parallel level and an incomplete parallel level of A}\scriptsize{T}\small{A-P, for $P=15$.} }
\label{fig: incomparlev}
\end{figure}

Hence, processes may be ordered depending on the call (HASA-P or A\footnotesize{T}\normalsize{A}-P) first, and on the size of the sub-problem as second parameter.  
In summary, the last $P - (k + 1) \cdot npl(\ell_{max})$ processes are paired, in a orderly way, to: (1) processes that are in charge of HASA-P calls, (2) processes that work on larger sub-problems. 
An example of how processes are distributed in a incomplete parallel level is depicted in Figure~\ref{fig: incomparlev}, for  $P = 15$. In this case, $\ell_{max} = 1$ and \emph{lefties}$= 9$. The value of $k$ resulting from equation (\ref{eq: k}) is 1, hence each of the $npl(\ell_{max})=6$ processes can be paired to one of the \emph{lefties} processes. As for the further remaining three processes, two are paired associated to processes performing HASA (namely, one is associated to the process having father with rank 4 and one is associated to the process having father with rank 5), whereas the third process must be associated to one of the process with higher sub-data size, that in this case is only $P_0$. 

\subsection{Implementation details}%Details on the MPI Implementation}
The parallel algorithm has been implemented using MPI for running on a multiprocessor system (see Section~\ref{sec:performance}). Once the MPI environment is initialized, every process gets its own identifying rank through the MPI function MPI\_rank. 
In addition to the input arguments given to  Algorithm~\ref{alg: AtA} in the sequential version, a parallel recursive call to  A\footnotesize{T}\normalsize{A}-P also reads: 
%\vspace{-\topsep}
\begin{itemize}
	\setlength{\parskip}{0pt}
	\setlength{\itemsep}{0pt plus 1pt}
	\item $\ell$ the index  of the level it is working on;
	\item $\ell_{max}$ the maximum number of complete parallel levels;
	\item \emph{father} the rank of the father of the process that is serving the current call;
	\item \emph{lefties} the number of processes beyond the ones needed to accomplish $\ell_{max}$ complete parallel levels. 
\end{itemize}
%\vspace{-\topsep}
%\begin{description}
%    \item[$\ell$] the index  of the level it is working on;
%    \item[$\ell_{max}$] the maximum number of complete parallel levels;
%    \item[\emph{father}] the rank of the father of the process that is serving the current call;
%    \item[\emph{lefties}] the number of processes beyond the ones needed to accomplish $\ell_{max}$ complete parallel levels. 
%\end{description}
The argument \emph{father} is the rank of the process that has called  A\footnotesize{T}\normalsize{A}-P, tracing the tree of the recursive calls (see Figure~\ref{fig: comparlev}). 

We assume that $P\geq 6$. When the function is first launched, the identifier of the generated parallel level is 1, and the father is process 0. At each parallel level of  A\footnotesize{T}\normalsize{A}-P, the sub-problems are divided among processes depending on their task. Initially, for each parallel recursive call, an array of size 6 called $ids$ is defined in the following way. In the  level $\ell$, the $i$-th element $ids_i$ of $ids$ is defined as $ids_i$ = \emph{father} +  $i\cdot npl(x)$, for $i = 0, 1,\ldots, 4$, while the 6-th element $ids_5$ is defined as $ids_5$ = \emph{father} + $4 \cdot npl(x) + 7^x$, where $x = \ell_{max} - \ell$. Afterwards, recursive calls  to complete parallel levels are assigned to processes with rank in [$ids_i, ids_{i+1}$) if $i < 5$, and to process with rank [$ids_5$, $npl(\ell_{max})$) otherwise, for working on different sub-problems. Notice that process $P_{ids_0}$ is $P_{father}$. Similarly, in each parallel level of HASA-P, the array $ids$ of size 7 is defined in such a way that the $i$-th element $ids_i$ of $ids$ is $ids_i$ = \emph{father} + $i\cdot 7^x$, where again $x = \ell_{max} - \ell$. It is easy to understand why the numbering is constructed in this way by looking at the tree shown in Figure~\ref{fig: comparlev}.

The pseudocode of a simplified version of  A\footnotesize{T}\normalsize{A}-P is reported in Algorithm~\ref{alg: AtAP}. 
This version of the algorithm is simplified in the sense that we assume that $P$ is taken equal to $npl(\ell_{max})$, for some $\ell_{max}$, and this implies that only complete parallel levels are executed. 
Since the input argument \emph{lefties} is necessary only when an incomplete parallel level can be performed, it is omitted in Algorithm~\ref{alg: AtAP}, (in fact, in this case \emph{lefties} is equal to 0, being $P=npl(\ell_{max})$).
This simplification can be removed simply by adding the management of the set of \emph{lefties} processes that we have when $P \neq npl(\ell_{max})$ and that is described in Section \ref{sec: ATAins}.

\subsection{Communication between  processes}
In the following sections \ref{sec: comATA} and \ref{sec: comHASA}, we shall describe how communication between processes was developed. For both functions A\footnotesize{T}\normalsize{A}-P and HASA-P, we combined communicators and point-to-point communication. Communicators group together processes that may be organized in different topology within the communicator they belong to. Point-to-point communication allows two specific processes to share data.  
%****** FIG 3 stava qui

\alglanguage{pseudocode}
\begin{algorithm}%[t]
\footnotesize
%\small
\label{alg: AtAP}
\caption{A\footnotesize{T}\normalsize{A}-P - simplified version: $P=npl(\ell_{max})$}
\label{alg: AtAP}
\textbf{Input:} $A\in \mathbb{R}^{m\times n}$ \\
\textbf{Output:} Lower triangular part of $C = A^t \cdot A$
\begin{algorithmic}[1]

\Procedure{AtA-P}{$A$, $m$, $n$, $\ell$, $\ell_{max}$, father}
\If {$m \lor n <=$ 32} \Return mult($A$, $m$, $n$); \hspace*{2.5cm}\%Base case
\Else \hspace*{8cm}\%Iterative case
\Statex \hspace*{8.72cm}\%Initialize sub-data
\State Define $m_1$, $n_1$, $m_2$, $n_2$;
\State Initialize $A_{i,j}$, $i, j = 1,2$;
\If{$\ell > 0$} \hspace*{6.4cm}\%$\ell>0$ characterizes a parallel level
\State $x \leftarrow \ell_{max} - \ell$;
\State $step \leftarrow npl(x)$;
\For{($i = 0; i < 4; i$++)}
\State $ids_i \leftarrow$ father + $i \cdot step$;
\EndFor
\State $ids_5\leftarrow$ father + $4 \cdot step + 7^x$;
\If{$\ell<\ell_{max}$} new\_$\ell \leftarrow \ell + 1;$
\Else \State new\_$\ell=0$;%new\_$\ell\leftarrow$UPDATE($\ell$, $\ell_{max}$, lefties);
\EndIf
\State $id\leftarrow$ MPI\_get\_rank;
%\State MPI\_get\_nprocs;
\Statex \hspace*{8.72cm}\%MPI: create communicators:
\State Comm$_{1,1} = \left \{ ids_0, ids_1\right \}$,
\State Comm$_{2,2} = \left \{ ids_2, ids_3\right \}$,
\State Comm$_{2,1} = \left \{ ids_4, ids_5\right \}$;
\State Current World = $\left \{ ids_0, \ldots, ids_5\right \}$;
\Statex \hspace*{8.72cm}\%Parallel recursion
\If{$id \in [ids_0, ids_1)$}

\State $S_{1,1}\leftarrow$ A\tiny{T}\footnotesize{A-P}($A_{1,1}$, $m_1$, $n_1$, new\_$\ell$, $\ell_{max}$, $ids_0$); 
\EndIf
\If{$id \in [ids_1, ids_2)$}
\State $S_{1,1}\leftarrow$ A\tiny{T}\footnotesize{A-P}($A_{2,1}$, $m_2$, $n_1$, new\_$\ell$, $\ell_{max}$, $ids_1$);  
\EndIf
\If{$id \in [ids_2, ids_3)$}
\State $S_{2,2}\leftarrow$ A\tiny{T}\footnotesize{A-P}($A_{1,2}$, $m_1$, $n_2$, new\_$\ell$, $\ell_{max}$, $ids_2$);  
\EndIf
\If{$id \in [ids_3, ids_4)$}
\State $S_{2,2}\leftarrow$ A\tiny{T}\footnotesize{A-P}($A_{2,2}$, $m_2$, $n_2$, new\_$\ell$, $\ell_{max}$, $ids_3$);  
\EndIf
\If{$id \in [ids_4, ids_5)$}
\State $S_{2,1}\parbox{.3cm}{\leftarrowfill}$\scriptsize{HASA-P}\footnotesize($A_{1,2}^t$, $A_{1,1}$, $n_2$, $m_1$, $n_1$, new\_$\ell$, $\ell_{max}$, $ids_4$); 
\EndIf
\If{$id \in [ids_5, npl(\ell_{max})$)}
\State $S_{2,1}\parbox{.3cm}{\leftarrowfill}$\scriptsize{HASA-P}\footnotesize($A_{2,2}^t$, $A_{2,1}$, $n_2$, $m_2$, $n_1$, new\_$\ell$, $\ell_{max}$, $ids_5$);  
\EndIf
\Statex \hspace*{8.72cm}\%Communication
\If{$id = ids_0 \lor id = ids_1$}  \State MPI\_Reduce(sendbuf: $S_{1,1}$, recvbuf: $C_{1,1}$, MPI\_op: MPI\_SUM, root:0, comm: $\text{Comm}_{1,1}$);
\EndIf
\If{$id = ids_2 \lor id = ids_3$}  \State MPI\_Reduce(sendbuf: $S_{2,2}$, recvbuf: $C_{2,2}$, MPI\_op: MPI\_SUM, root:0,
comm: $\text{Comm}_{2,2}$);
\EndIf
\If{$id = ids_4 \lor id = ids_5$}  \State MPI\_Reduce(sendbuf: $S_{2,1}$, recvbuf: $C_{2,1}$, MPI\_op: MPI\_SUM, root:0, comm: $\text{Comm}_{2,1}$);
\EndIf
\If{$id = ids_2$} Send $C_{2,2}$ to $P_{ids_0}$;
\EndIf
\If{$id = ids_4$} Send $C_{2,1}$ to $P_{ids_0}$;
\EndIf
\If{$id = ids_0$} Receive $C_{2,2}$ from $P_{ids_2}$ and $C_{2,1}$ from $P_{ids_4}$;
\State \Return  = \[ C
\begin{bmatrix}

C_{1,1} &  \\
C_{2,1} & C_{2,2} \\
\end{bmatrix} 
\]
\Else  \State return NULL;
\EndIf
\Else \If{$\ell = 0$}  \hspace*{6.15cm}\%When new parallel levels cannot be 
\State \Return A\tiny{T}\footnotesize{A}(A, m, n);\hspace*{4.8cm}\%generated, processes go sequentially
\EndIf
\EndIf
\EndIf
\EndProcedure
\end{algorithmic}
\end{algorithm}

\subsubsection{Communication in A\footnotesize{T}\normalsize{A}-P}
\label{sec: comATA}
In this section we describe how communication is carried out in A\footnotesize{T}\normalsize{A}-P complete parallel levels. In each complete parallel level, the six processes generated by the call to A\footnotesize{T}\normalsize{A}-P are responsible for the four recursive calls to A\footnotesize{T}\normalsize{A}-P and for the two calls to HASA-P, and are now identified by rank $ids_0,\ldots, ids_5$ (where $ids$ is the array defined as in the previous section). Recall that, for each call, $ids_0$ is the father of itself and of the remaining elements of $ids$. Tasks are divided in the following way:\\
\newline
$P_{ids_0}$ runs A\footnotesize{T}\normalsize{A}-P  on $A_{1,1}\rightarrow S_{1,1}=A_{1,1}^t\cdot A_{1,1}$;\\
$P_{ids_1}$ runs A\footnotesize{T}\normalsize{A}-P  on $A_{2,1}\rightarrow S_{1,1}=A_{2,1}^t\cdot A_{2,1}$;\\
$P_{ids_2}$ runs A\footnotesize{T}\normalsize{A}-P on $A_{1,2}\rightarrow S_{2,2}=A_{1,2}^t\cdot A_{1,2}$;\\
$P_{ids_3}$ runs A\footnotesize{T}\normalsize{A}-P  on $A_{2,2};\rightarrow S_{2,2}=A_{2,2}^t\cdot A_{2,2}$;\\
$P_{ids_4}$ runs HASA-P on ($A_{1,2}^t$, $A_{1,1})\rightarrow S_{2,1}=A_{1,2}^t\cdot A_{1,1}$;\\
$P_{ids_5}$ runs HASA-P on ($A_{2,2}^t$, $A_{2,1}) \rightarrow S_{2,1}=A_{2,2}^t\cdot A_{2,1}$.\\

\begin{figure*}[t]
\centering
\includegraphics[width=0.98\textwidth]{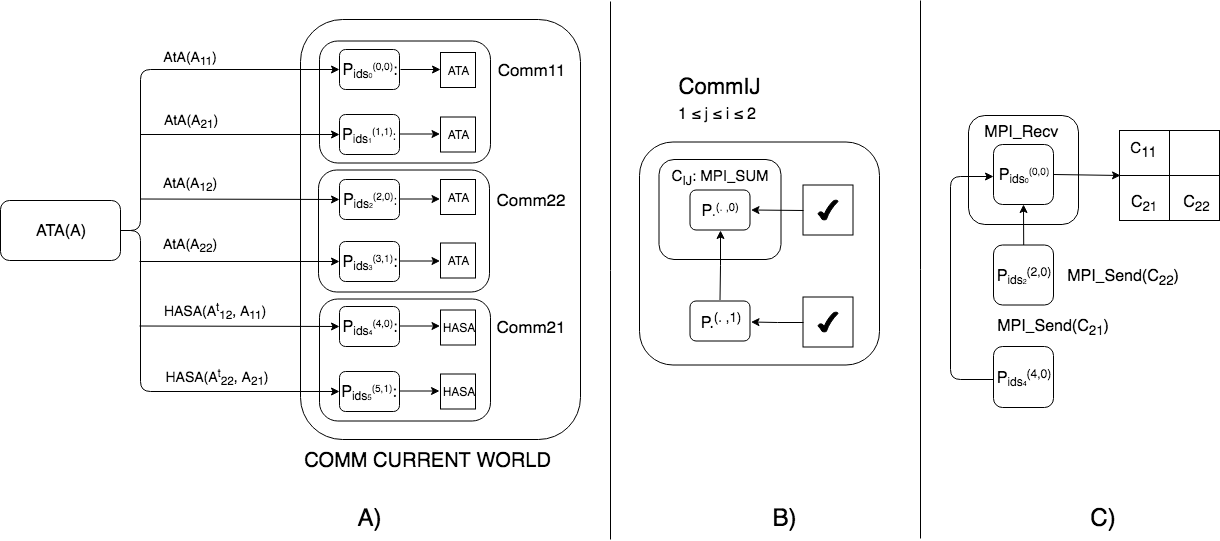}
\caption{\small Representation of the three steps of communication in a complete parallel level of A\scriptsize{T}\small{A}-P. 
\hspace{0.15cm} A) A\scriptsize{T}\small{A}-P is executed on matrix $A$. The communicator Current World collects together the six processes involved. Communicators Comm11, Comm22 and Comm21 group together the processes that compute the addends of $C_{1,1}$, $C_{2,2}$ and $C_{2,1}$, respectively, where $C=A^tA$. Process denoted as $P_{ids_i}^{(k,l)}$ is the process with rank $ids_i$ in COMM\_WORLD, $k$ in Current World and $l$ in the smaller communicator CommIJ.
\hspace{0.15cm} B) When the recursion is over, a MPI reduction computing a matrix sum is performed within each communicator CommIJ. The result is stored in the process $P^0$. 
\hspace{0.15cm} C) Processes of rank 2 and 4 in Current World send $C_{2,2}$ and $C_{2,1}$ respectively to $P_0$, that is now in charge for patching together and returning $C$.}
\label{fig: commAtA}
\end{figure*}

Each of the three pairs of processes ($P_{ids_0}$,$P_{ids_1}$), ($P_{ids_2}$, $P_{ids_3}$), and ($P_{ids_4}$, $P_{ids_5}$) is responsible for computing  the two addend of the three sub-matrices of $C=A^t\cdot A$ (see equation (\ref{eq: Cij})).
Four communicators are created at this stage. 
Three communicators link together the addends of each of the three sub-matrices $C_{1,1}$, $C_{2,2}$ and $C_{2,1}$. 
To this end, an MPI reduction performing a matrix sum is executed within each of such communicators. As a result, processes $P_{ids_0}$, $P_{ids_2}$ and $P_{ids_4}$ have $C_{1,1}$, $C_{2,2}$ and $C_{2,1}$ respectively. 
The fourth communicator, that we call Current World, collects together all processors in $ids$.
Current World is used to handle MPI barriers to synchronize processes, guaranteeing safe communication. By convention, the root of communication is the process with the lowest rank in the current communicator. Blocking Send and Receive functions are employed to transfer $C_{2,2}$ and $C_{2,1}$ from  $P_{ids_2}$ and $P_{ids_4}$, respectively, to processor $P_{ids_0}$, which is now in charge of defining and returning $C$. Communication is represented thoroughly in Figure \ref{fig: commAtA}.

\subsubsection{Communication in HASA-P}
\label{sec: comHASA}
Here below we show how we implemented the communication system in a HASA-P complete parallel level. In each complete parallel level, the seven processes with rank $ids_0,\ldots, ids_6$ are responsible for the seven recursive calls to HASA-P. 
In particular if HASA is run on matrices $A$ and $B$, to obtain their product $D = A\cdot B$, the workload is distributed among processes as follows:\\
\newline
$P_{ids_2}$ runs HASA on ($A_{1,1}$, $B_{1,2} - B_{2,2})\rightarrow M_3$;\\
$P_{ids_5}$ runs HASA on ($A_{2,1} - A_{1,1}$, $B_{1,1} + B_{1,2})\rightarrow M_6$;\\
$P_{ids_1}$ runs HASA on ($A_{2,1} + A_{2,2}$, $B_{1,1}) \rightarrow M_2$;\\
$P_{ids_0}$ runs HASA on ($A_{1,1} + A_{1,2}$, $B_{1,1} + B_{2,2})\rightarrow M_1;$\\
$P_{ids_4}$ runs HASA on ($A_{1,1} + A_{1,2}$, $B_{2,2})\rightarrow M_5$;\\
$P_{ids_6}$ runs HASA on ($A_{1,1} - A_{2,2}$, $B_{2,1} - B_{2,2})\rightarrow M_7$;\\
$P_{ids_3}$ runs HASA on ($A_{2,2}$, $B_{2,1} - B_{1,1})\rightarrow M_4$.\\
\newline
When processes $P_{ids_0},\ldots, P_{ids_6}$ complete their work, the product matrix $D=A\cdot B$ is obtained calculating its four blocks as follows:\\
\newline
$D_{1,1}=M_1 - M_5 +M_7 +M_4$;\\
$D_{1,2} = M_3+M_5;$\\
$D_{2,1} = M_2+M_4;$\\
$D_{2,2} =M_3+M_6 -M_2 + M_1$.
\newline

\begin{figure}[t]
\centering
\includegraphics[width=0.45\textwidth]{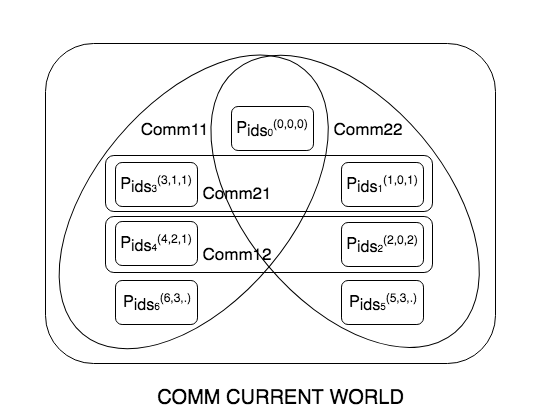}
\caption{\small Representation of communicators in a HASA-P parallel level step. $P_{ids_i}^{(k,l,h)}$ is process with rank $ids_i$ in COMM\_WORLD, $k$ in Current World, $l$ (and possibly $h$) in the first and second smaller communicators CommIJ.} 
\label{fig: commHASA}
\end{figure}

Similarly to how was done for  A\footnotesize{T}\normalsize{A}-P, four communicators (one for each block of $D$) are created at this stage. 
Processes computing a term that appears in more than one block of $D$ (namely, all terms but $M_6$ and $M_7$) belong to two communicators. A MPI reduction allows to store $D_{i,j}$, $i, j = 1,2$ in the root of each communicator. Once all MPI reductions are completed, all data is sent to $P_{ids_0}$ using Send/Receive functions. $P_{ids_0}$ is responsible for recovering and returning $D$. 
Similarly to what described for A\footnotesize{T}\normalsize{A}-P, synchronization among processes is achieved by creating a Current World communicator that includes all processes $P_{ids_i}$, $i = 0,\ldots, 6$. 
Communicators are represented in Figure \ref{fig: commHASA}.

%Some processes belong to two communicators, as the sub-matrix $M_i$ that they compute appears as a term of more than one block of $D$. 

\section{Communication model and cost}
\label{sec:comm-cost}
Our communication model is similar to the one used in \cite{ballard2012communication}. We consider latency and bandwidth costs, denoted as $L(n, P)$ and $BW(n, P)$, respectively. Latency cost is the communicated message count while bandwidth cost is expressed in terms of communicated word count. Message and word counts are computed along the critical path introduced in \cite{yang1988critical}. If $\alpha$ is the time spent for communicating a message and $\beta$ is the time for communicating a word, then the total communication cost is given by: 
\begin{equation*}
    \alpha L(n, P) + \beta BW(n, P).
\end{equation*}
The number of messages that are exchanged between processes depends on how many complete parallel levels of processes can be layered, $\ell_{max}$ (see equation (\ref{eq: lmax})). 
Notice that $\forall P$ it holds $\ell_{max}<\log_7 P$; this is because of the number of nodes of the ideal tree that processes are distributed on (see Figure \ref{fig: comparlev} for an example); more precisely, we may observe that for all $P\leq 11602$  it holds that $\ell_{max} = O(\log_{6.5}P)$.

In a A\footnotesize{T}\normalsize{A}-P communication step, three MPI reductions occurs simultaneously. Afterwards, two messages are sent to the root of the Current World communicator (see Figures \ref{fig: commAtA} B) and C)). In a HASA-P communication step, four MPI reductions are performed at the same time, and three messages are sent to the Current World root. Since HASA-P is recalled on $\ell_{max} -1$ levels, it holds that $L(n,P) = \max \left \{ 4\cdot (\ell_{max} - 1); 3\cdot \ell_{max} \right \}$, that is $L(n, P) = \max \left \{ 4\cdot (O(\log_7P) - 1); 3\cdot O(\log_7 P) \right \}$ in general. 
%Despite 
Hence, we can observe that the achieved latency is lower than the one reached in \cite{ballard2012communication} for generic $A\cdot B$ multiplication, but communication suffers from high bandwidth cost ($BW(n, P) = \big ( \frac{n}{2} \big )^2$). %We shall comment 
Insights on this fact are given in Section \ref{sec: imp2}. %that influences performances, as we shall see in the next section.

\section{Performance evaluation}
\label{sec:performance}
In this section we describe the results of experimental tests carried out in order to assess A\footnotesize{T}\normalsize{A}-P. We implemented A\footnotesize{T}\normalsize{A}-P using MPI and tested it on a cluster of Intel processors. 
Performances are analyzed using different number of processors and several features for parallel algorithms were investigated.  
\subsection{Experimental setup}
Performances have been tested on a small subset of nodes of the Galileo cluster, installed in CINECA (Bologna, Italy), \cite{galileo}.
It is an IBM NeXtScale, Linux Infiniband Cluster consisting of 360 nodes 2 x 18-cores Intel Xeon E5-2697 v4 (Broadwell) processors (2.30 GHz), and 15 nodes 2 x 8-cores Intel Haswell (2.40 Ghz ) processors endowed with 2 nVidia K80 GPUs.

\subsection{Experimental results}
We studied the performances of A\footnotesize{T}\normalsize{A}-P in terms of execution time, speed-up, efficiency and Karp-Flatt metric. For tests, we considered randomly generated square matrices of size $n = 5000$ and $n = 10000$. Tests have been carried out for the following values of activated processes $P= 6, 12, 18, 38, 76, 114, 250$. The cases $P=6, 38, 250$ correspond to a parallel execution with complete parallel levels, whereas the remaining values of $P$ generate incomplete parallel levels.

\begin{figure}[h]
\centering
\includegraphics[width=0.5\textwidth]{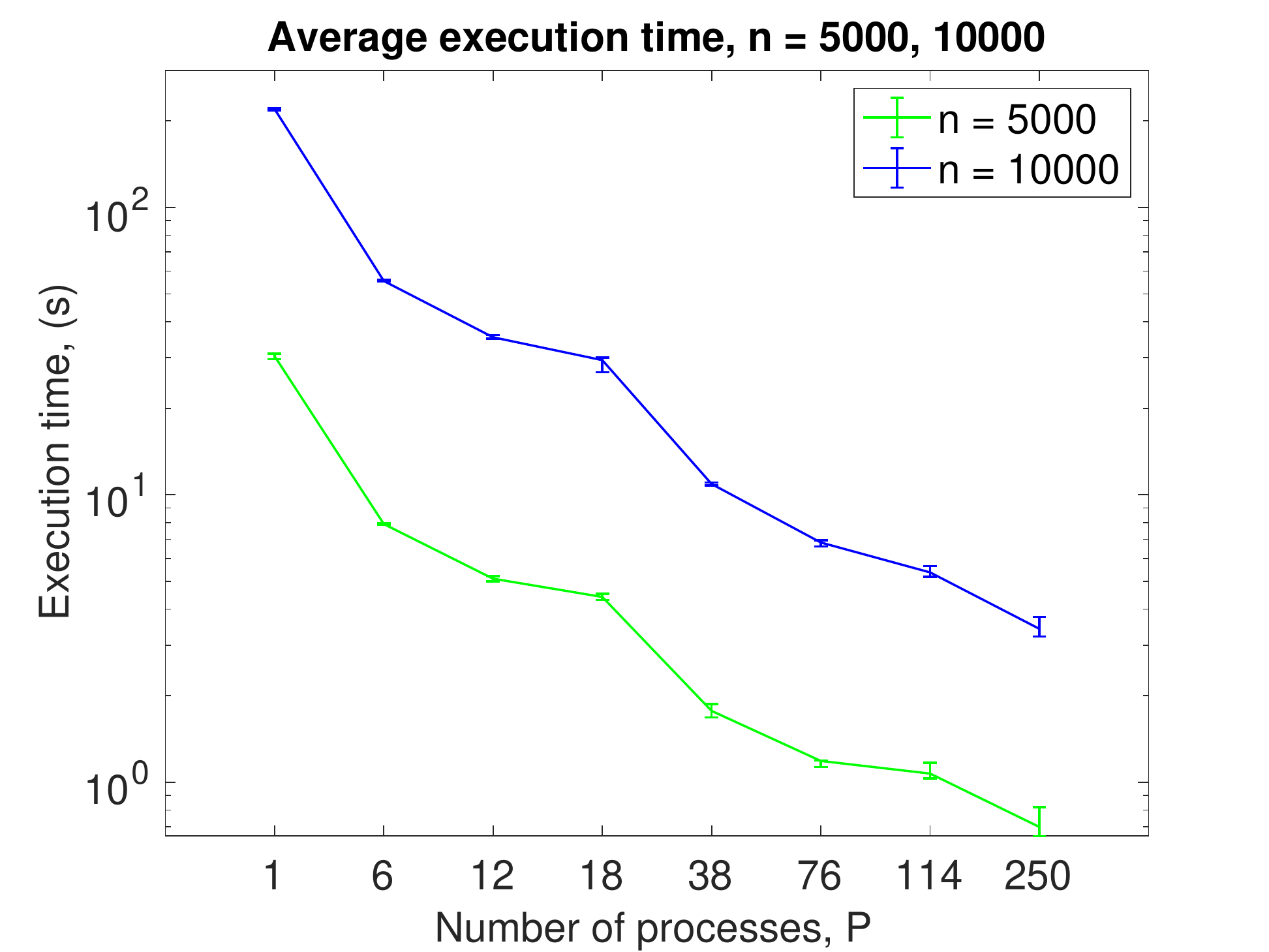}
\caption{\small Execution time for  A\scriptsize{T}\small{A}-P for matrix sizes 5000 and 10000, varying the number of processes. value for $P=1$ represents serial execution time. }
\label{fig: ExecTime}
\end{figure}

\begin{figure}[h]
\centering
\includegraphics[width=0.5\textwidth]{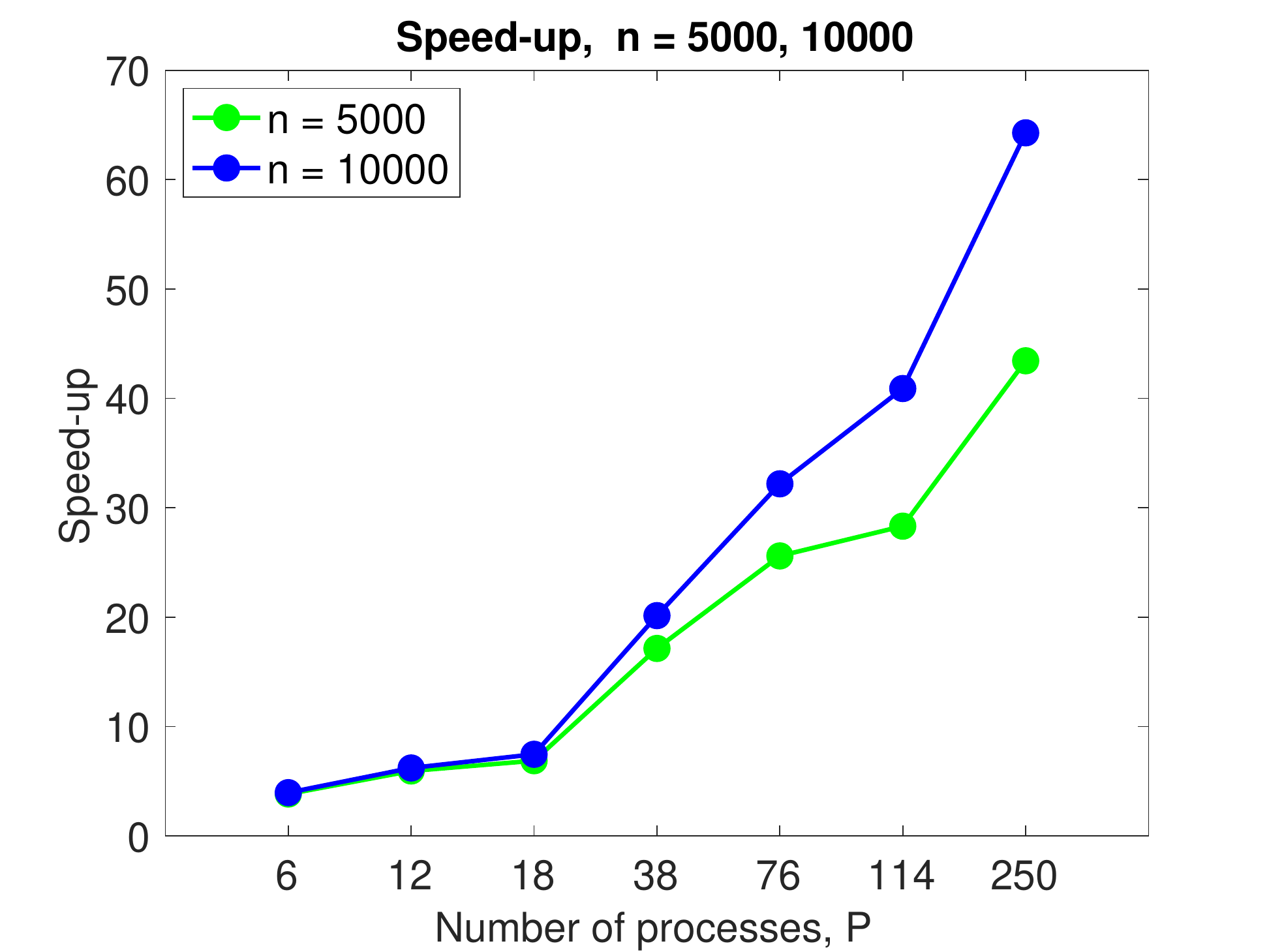}
\caption{A\scriptsize{T}\normalsize{A}-P speed-up, matrix size = 5000, 10000.  }
\label{fig: speedup}
\end{figure}

Execution times for matrix size $5000$ and $10000$ are depicted in Figure~\ref{fig: ExecTime}. For $P=1$, the execution time is obtained running the sequential implementation of A\footnotesize{T}\normalsize{A}. 
From Figure~\ref{fig: ExecTime}, we can observe that for values of $P$ that do not correspond to the activation of complete parallel levels there is a degradation in the trend of the execution time. On the contrary when the number of processes allows to generate complete parallel levels, the execution time improves.
This aspect is also observable in Figure~\ref{fig: speedup}, where we show the speed-up values.
%, but it is clearly evident in Figure~\ref{fig: efficiency}, where the values of the efficiency are reported.  
We can observe that for matrices of size $n=10000$ the speed-up maximum value is $64.28\%$, and is obtained for $P=250$, see Figure~\ref{fig: speedup}. 
The trend of the execution times shown in Figure~\ref{fig: ExecTime} is strictly decreasing, highlighting the good scalability of the approach. %a further potential scalability for higher values of $P$. Speed-up increases accordingly.

\begin{figure}[h]
\centering
\includegraphics[width=0.5\textwidth]{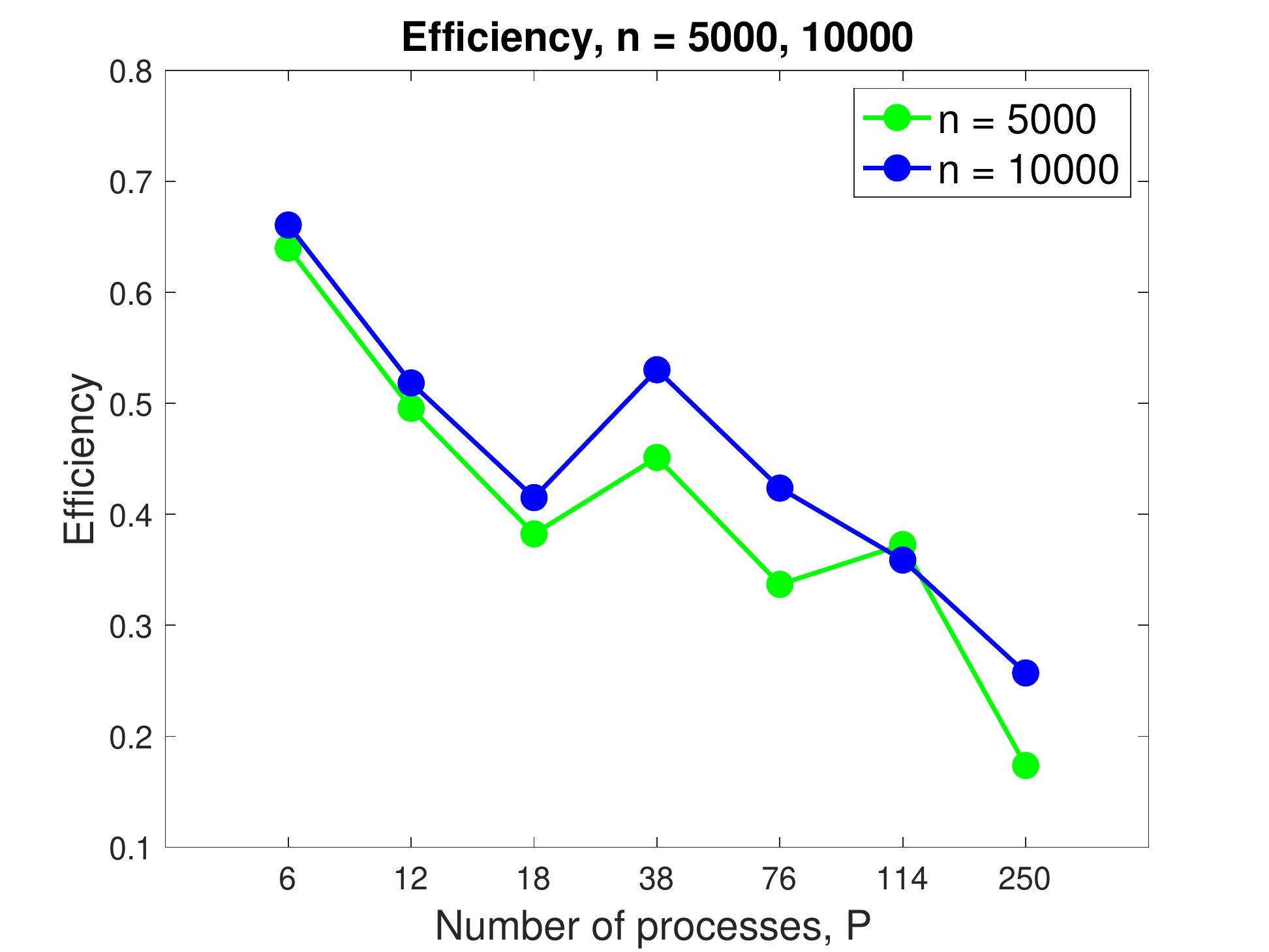}
\caption{\small A\scriptsize{T}\small{A}-P efficiency, matrix size = 5000, 10000. }
\label{fig: efficiency}
\end{figure}

The efficiency is computed as the ratio between the speed-up $S$ and the number of processes $P$, and ranges from 0.66 (obtained for $P=6$) to 0.26 (obtained for $P=250$), as reported in Figure~\ref{fig: efficiency}. 
As expected, efficiency has a decreasing trend, except for $P=38$, where it grows. A first observation about this fact is the following: for $P=38$, exactly 2 complete parallel levels can be performed, while for $P=18$ it holds that $\ell_{max}=1$ and the remaining 12 processes are paired to the $npl(1)=6$ processes in order to create an incomplete parallel level. Therefore, some portions of executed code are not shared in the two cases arising from having $P=18$ and $P=38$. Further comments on this phenomenon are in Section~\ref{sec: imp2}. 

\begin{figure}[h]
\centering
\includegraphics[width=0.5\textwidth]{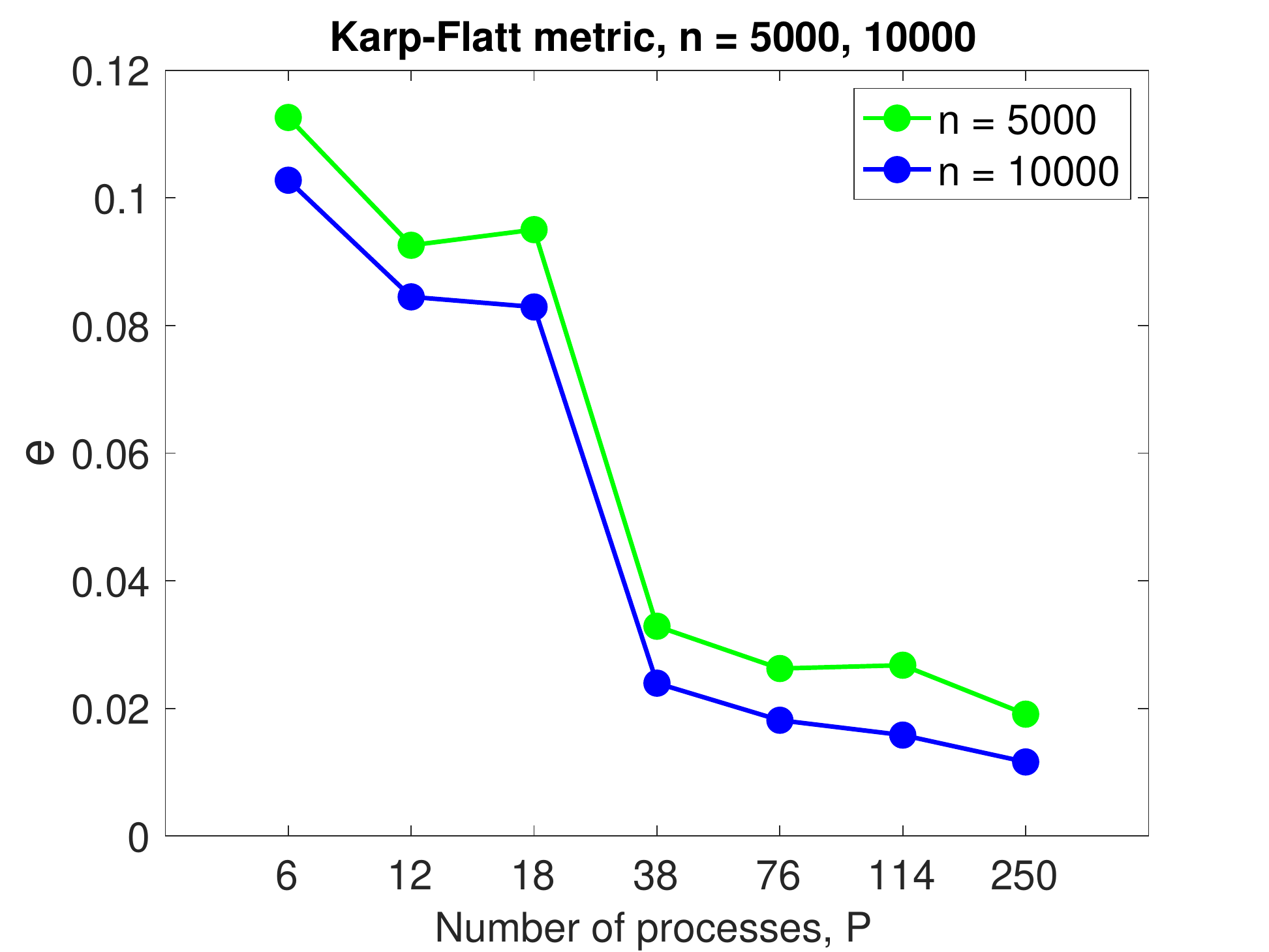}
\caption{\small Karp-Flatt metric, matrix size = 5000, 10000.}
\label{fig: Karp-Flatt}
\end{figure}

Finally, in Figure~\ref{fig: Karp-Flatt}, the Karp-Flatt metric values, giving the \emph{experimentally determined serial fraction}, are reported. 
The Karp-Flatt metric was first introduced in \cite{karp1990measuring} and depicts the fraction of time spent by a parallel program to perform serial code, $e$. It is defined as follows:
\begin{equation}
    e = \frac{\frac{1}{S} - \frac{1}{P}}{1 - \frac{1}{P}},
\end{equation}
where $S$ is the speed-up and $P$ is the number of processes. 
As we can see in Figure \ref{fig: Karp-Flatt}, values of $e$ are small and decreasing, meaning that there is no significant parallel overhead and that the portion of serial code that is executed is very low. %\textcolor{green}{Notice that, for all parameters, performance improves slightly for matrix size = 10000. }

%\subsection{Performance improvement}
\subsection{Discussion of performance results}
The performance assessment parameters show that A\footnotesize{T}\normalsize{A} is scalable and highlight high speed-up and negligible portions of executed serial code.  Efficiency has a frequently observed decreasing trend that is due to intrinsic parallel overhead when the number of processes $P$ grows. In this section we make two key observations that may be taken into account for improving the performances of A\footnotesize{T}\normalsize{A}-P. 

%\subsubsection{Improvement 1: Process redistribution}
\subsubsection{Process redistribution}
As we described in Section \ref{sec: par_imp}, every time A\footnotesize{T}\normalsize{A}-P and HASA-P are called recursively, in each complete parallel level one process is responsible for one recursive call, either to A\footnotesize{T}\normalsize{A}-P or to HASA-P. Nevertheless, the computational cost of HASA-P is higher than the one of A\footnotesize{T}\normalsize{A}-P. This results in a idle time for those processes performing A\footnotesize{T}\normalsize{A}-P that worsen efficiency. As a matter of fact, the peak of efficiency for $P=38$ (see Figure \ref{fig: efficiency}) is rather due to a low efficiency for $P=12$ and $P=18$, that are the two configurations that report the highest time difference between processes executing  A\footnotesize{T}\normalsize{A}-P and those performing HASA-P.
A strategy to overcome this issue can be to distribute processes so that more processes are responsible for HASA-P calls, at the expenses of more workload for  those performing A\footnotesize{T}\normalsize{A}-P. 

%\subsubsection{Improvement 2: Data redistribution}
\subsubsection{Data redistribution}
\label{sec: imp2}
In Section \ref{sec:comm-cost} we discussed the expression for latency and bandwidth of A\footnotesize{T}\normalsize{A}-P. We noticed that the number of sent messages is small, but that the maximum size of sent messages is independent from $P$. Because of the low latency, this does not introduce a very appreciable delay when the number of processes is low, but it may introduce non negligible overhead for a very high number of processes. In the configurations that we investigated, the maximum percentage of time spent for communication ranges between $\sim$0.14\% ($P=6$, maximum time spent for communication is 0.08s) and $\sim$0.46\% ($P=250$, maximum time spent for communication is 0.16s) of the total parallel execution time. A possible solution to high bandwidth is to divide matrix size such that all processes work on the same amount of data and avoiding intermediate communication between processors.

\section{Conclusions and future work}
\label{sec:conclusions}
We have defined a cache-oblivious recursive algorithm for the $A^tA$ matrix multiplication, A\footnotesize{T}\normalsize{A}, that is an operation that has applications in several problems in geometry, linear algebra, statistics, etc. The number of multiplications performed on matrices of size $n$ is  upper-bounded by $\frac{2}{7}n^{\log_2 7}$, in the face of $n^2(n+1)/2$ products for conventional $A^tA$ multiplication algorithm; this is achieved because A\footnotesize{T}\normalsize{A} includes recursive calls to generalized Strassen's algorithm for rectangular matrix multiplications. The algorithm was implemented in parallel using MPI and tested on a cluster. MPI communication facilities were used to perform smart communication in each parallel execution of A\footnotesize{T}\normalsize{A}. Latency is low and the performances were assessed in terms of several performance evaluation indices, highlighting good scalability, low parallel overhead and negligible fraction of inherently sequential code. We detected possible improvements for the enhancement of parallel performances consisting in a different balance for task distribution among processes, and a more equally spread load of data among processors. We plan to find a trade-off between the two proposed improvements. Also, we believe that existing performance evaluation indices penalize the test results of systems where the execution time of employed processors do not overlap perfectly: first and foremost, conventional efficiency does not take into account the amount of time during which not all $P$ processors are actively working; yet processes performing faster operations may use idle time for fulfilling additional tasks if the algorithm is integrated in a more complex system. We believe that a more accurate formulation for efficiency and speed-up may be useful to assess more truthfully systems like the one that we introduced. 

\section*{Acknowledgements}
This work has been partially supported by MIUR grant Excellence Departments
2018-2022, assigned to the Computer Science Department of Sapienza University of Rome.
The experimental part has been run on the Galileo cluster, located in Cineca, thanks to Class C ISCRA Project n. HP10CCM8RG.

%\clearpage
%\newpage
%\bibliographystyle{plain}
\bibliographystyle{plain}
\bibliography{references}
\end{document}